\documentstyle[12pt]{article}
\input amssym.def
\input amssym
\def \dst {\displaystyle}

\newcommand{\la} {\lambda}

\newcommand{\af} {\hat{A}}

\newcommand{\al} {\alpha}

\newcommand{\be} {\beta}
\newcommand{\ga} {\gamma}

\newcommand{\ffi}{\varphi ( \ga, \eta )}
\newcommand{\ffa}{\varphi ( \al, \be )}
\newcommand{\ffn}{\hat{\varphi} ( \alpha , \eta )}
\newcommand{\bga}{\begin{array}{l}}
\newcommand{\ena}{\end{array}}
\newcommand{\bge}{\begin{equation}}
\newcommand{\ene}{\end{equation}}
\title{ Light-Ray Radon Transform for  Abelianin and Nonabelian
Connection in 3 and 4 Dimensional Space with Minkowsky Metric .}
\author{ M. Zyskin\thanks{Department of Physics and Astronomy,
Rutgers
University, Piscataway, NJ 08855}}
\date{}
\begin{document}
\maketitle
\begin{abstract}
\vbox{
\noindent
 We have a real manifold of dimension 3 or 4  with Minkovsky
metric, and with a connection for a trivial  $GL(n, C)$ bundle  over that
manifold. To each light ray on the manifold we assign  the data of paralel
transport along that light ray. It turns out that these data are not enough to
reconstruct the connection, but we can add more data, which depend now not from
lines but from 2-planes, and which in some sence are the data of parallel
transport in the complex light-like directions, then we can reconstruct the
connection up to
a gauge transformation. There are some interesting applications of the
construction: 1) in 4 dimensions, the self-dual Yang Mills equations
can be written as the zero curvature condition for a pair of certain first
order differential operators; one of the operators in the pair is the covariant
derivative in complex light-like direction we studied.\\
2) there is a relation of this Radon transform with the
supersymmetry.\\
3)using our Radon transform, we can get a measure on the space of 2 dimensional
planes in 4 dimensional real space. Any such  measure give rise to a Crofton
2-density. The integrals of this 2-density over surfaces in ${\Bbb R^4}$ give
rise to the Lagrangian for maps of real surfaces into ${\Bbb R^4}$, and
therefore to  some string theory.\\
4) there are relations with the representation theory. In particular,
a closely related transform in 3 dimensions can be used to get the Plancerel
formula for representations of $SL(2,R)$.}
\end{abstract}

\newpage
\setcounter{section}{0}
\section{Introduction .}

Let $M$ be the real 3 or 4 dimensional manifold with Minkowski metric.
Let us take some connection $\Omega=A_m(x)dx^m$. Consider
the paralel transport
equation in the light-like direction  :
$$
\al^m \left(\dst\frac{\partial}{\partial x^m} - A_m (x) \right)\mu(\al , x) =0,
$$
where
$$
\al_m \al^m = 0.
$$
Here $x\in M$;  $A_m(x)$ and  $ \mu(\al,x)$ in the abelian
case are with values in numbers, and in the nonabelian
case have  values
in complex $n\times n $ matrices. (For simplicity, we consider
the $gl(n,C)$ case only.)
\vskip 15pt

\noindent {\em Remark.} We use the summation convention: summation
over repeated
indexes is always assumed.
\vskip 15pt

\noindent $\mu(\al,x)$   depends on the point and on the ray we choose.
We will consider
certain  functionals of $\mu$, which depend only on the light ray. We call
such functionals spectral data of the light-ray Radon transform.
An example of spectral data functional: take the asymptotic of
$\mu(\al,x)$  along the light ray
$x=\al t + \be$ as $t\rightarrow -\infty$ to be 1, and
compute $\varphi ( \alpha , \beta ):= {lim}_{t \rightarrow +\infty}
\mu(\al,\al t + \be)$.

The spectral data are functionals of the connection $\Omega$.
For a  given connection $\Omega$, the spectral data are functions on
the space of light rays. We will show that it is possible to choose such
spectral data functionals that the inverse functional exist, namely, if we
are given the values of the spectral data for every light ray, we are able
to reconstruct the connection up to a gauge transformation
$A_m (x)\rightarrow g(x)
A_{m}(x)g^{-1}(x)-g(x)\dst\frac{\partial}{\partial x^m} g^{-1}(x)$ \
(see Lemma 1.3 and Lemma 2.2). In dimension 4, the space of all light
rays has dimension
5, therefore, there is a compatibility condition for scattering data.

The problem to describe the scattering data and to write
inversion formulas was suggested by prof. E. Witten.

The motivation to study the light-ray Radon transform came from
supersymmetric Yang-Mills theories. It is known \cite{w} that in
certain sence supersymmetry transformation is  a
square root from light-like translations. Namely, let $Q_\al,
{\bar{Q}}_{\dot{\al}}$ be the generators of the supersymmetry algebra
$$
\bga
\{Q_\al,Q_\be\} = \{ {\bar{Q}}_{\dot{\al}} , {\bar{Q}}_{\dot{\be}}\}= 0
\\[14pt]
\{Q_\al,{\bar{Q}}_{\dot{\be}}\}= 2{\sigma^m}_{\al, \dot{\be}}\
\ i\frac{\partial}{\partial x_m}.
\ena
$$
where ${\sigma^m}_{\al, \dot{\al}}$ are Pauli matrices. Take
complex numbers ${\la}^{\al}, \eta^{\dot{\al}}$ . Then
$$
\frac{1}{2} {\left({\la}^\al Q_\al + \eta^{\dot{\al}}
{\bar{Q}}_{\dot{\al}}\right)}^2 = {\la}^\al \eta^{\dot{\al}}
{\sigma^m}_{\al, \dot{\al}}  \ i\frac{\partial}{\partial x_m}
$$
is a light-like translation: take $v^m =  {\la}^\al \eta^{\dot{\al}}
{\sigma^m}_{\al, \dot{\al}} $; then $v^m v_m =0$.
If $\eta^{\dot{\al}}=\bar{\la}^\al$,
(where bar means complex conjugate), then $v^m$ is real.

In recent solutions for supersymmetric Yang-Mills theories \cite{sw} the
Yang-Mills connection was written as Gauss-Manin connection, using the known
monodromies on an auxiliary complex plane. In our paper, we have a similar
story for
nonsupersymmetric gauge theories, namely, we express the nonsupersymmetric
Yang-Mills connection through the solution of some auxiliary Riemann-Hilbert
problem on complex plane.

The techique, similar to the technique developed in this paper, is useful for
problems in real integral geometry. Also, we expect some modification of these
technique will allow to obtain the Plancherel-type formulas for representation
of $SL(2,R)$, using methods of integral geometry. Details will be published
elsewhere.

\section{Spectral data, associated with parallel transport in the
light-like direction in dimension 3.}
\setcounter{subsection}{-1}
\subsection{The transform for the function}
Consider the  {\bf standart} Radon transform of the function, that is the
integral of the function over 2 dimensional planes:
$$
F(\omega_0, \omega_1, \omega_2;p):=\dst\int \delta(\omega x - p) f(x) d^3 x.
$$
This transformation has the inverse:
$$
f(x)=\dst\int \delta^{\bf \prime\prime}(\omega x - p) F(\omega,p) d_S
\omega dp
$$
where  $\omega$ is integrated over the unit sphere.

A plane with ${\omega_1}^2 + {\omega_2}^2 - {\omega_3}^2 >0$ contains two
1-parametric families of  parallel light rays, belonging to such plane,
thus the integrals of the function over such plane can be written through
the integrals of the function over the light rays.  Consider a light ray
$x=\al t + \be$, where ${\al_1}^2 + {\al_2}^2={\al_0}^2$. Let $\varphi
(\al,\be)$ be the integral of the function over the light ray,
$$
\varphi (\al, \be)= \dst\int f(\al t + \be) dt.
$$

\noindent Then
$$
F(-\al_1\omega_1-\al_2 \omega_2,\ \omega_1,\omega_2;\ p) = \dst\int \varphi
(1,\al_1, \al_2; \ 0,\be_1,\be_2) \ \delta (\be_1 \omega_1+\be_2 \omega_2-p) d
\be_1 d\be_2
$$
Thus, in the inversion formula the integrals over 2 dimensional planes with
${\omega_1}^2 + {\omega_2}^2 - {\omega_3}^2 >0$ can be expressed through the
integrals over light rays, and the other planes we need to keep.  The integrals
over the remaining planes are related to the  {\bf complex} light rays in the
following way: for real light rays,
for pairs $\al, \eta$ such that
\bge
\bga
{(\al^1)}^2 + {(\al^2)}^2-{(\al^0)}^2 =0
\\[14pt]
\eta_0 \al^0 + \eta_1 \al^1 + \eta_2 \al^2 =0.
\ena
\ene
at fixed $\al$, the function $\varphi (\al, \be) \ e^{i \be^m \eta_m}$, as
function of $\be$, does not change if we add to $\be$ a vector proportional to
$\al$, thus it can be restricted to the factor-space of ${\bf R^3}$ over the
1-dimensional subspace spanned by $\al$. We define
$$
\Phi(\al, \eta) = \dst\int_{{\bf R^3}/ \{ \al\}} \varphi (\al, \be) \
e^{i \be^m \eta_m} d_{\al} \be
$$
where the integral is over the factor space ${\bf R^3}/ \{\al\}$, and
$ d_{\al} \be =\vert \al^0 d\be^1\wedge d\be^2 + \al^1 d\be^2\wedge
d\be^0+\al^2 d\be^0 \wedge d\be^1\vert$  is a nonoriented volume
element on this space. It's easy to see that
$\Phi (\al, \eta)= \hat{f}(\eta)$, where $\hat{f}(\eta)$  is the Fourier
transform of $f(x)$. Thus, we have the following :
$$
\bga
-{(\al^0)}^2+ {(\al^1)}^2+ {(\al^2)}^2 =0
\\[14pt]
\eta_0 \al^0 + \eta_1 \al^1+ \eta_2\al^2=0
\\[14pt]
\Phi (\al, \eta)=\hat{f}(\eta)
\ena
$$
For real $\al$, it follows that ${\eta_1}^2 + {\eta_2}^2 \geq {\eta_0}^2$,
otherwise there are no solutions. Thus we take complex $\al$ and consider
this system as the definition of what $\Phi (\al, \eta)$ is. Take $\al
=(1,\dst\frac{\la+\la^{-1}}{2},-i \dst\frac{\la-\la^{-1}}{2})$. Solving for
$\eta_1, \eta_2$ we will have
$\Phi(\al, \eta)=\hat{f}( \eta_0, -  \dst\frac{\la + \bar{\la}}{1+\la
\bar{\la}}\eta_0 , i \dst\frac{\la - \bar{\la}}{1+\la \bar{\la}}\eta_0)$.
After the Fourier transform over the remaining $\eta_0$ we get
$$
\dst\int \hat{f}(\eta_0, -  \dst\frac{\la + \bar{\la}}{1+\la \bar{\la}}\eta_0 ,
i \dst\frac{\la - \bar{\la}}{1+\la \bar{\la}}\eta_0)e^{i\eta_0 p}\dst\frac{ d
\eta_0}{2\pi}= \dst\int \delta( x^0 -  \dst\frac{\la + \bar{\la}}{1+\la
\bar{\la}} x^1 +  i \dst\frac{\la - \bar{\la}}{1+\la \bar{\la}}x^2 -p) f(x) d^3
x
$$
For $|\la|\neq 1$ this is the integral over those 2-planes
which do not contain real light rays, and as we change $\la$, we spann all of
them. Notice
also, that
in some sence the plane above contain the complex light ray $(1,\dst\frac{\la +
\la^{-1}}{2},- i \dst\frac{\la - \la^{-1}}{2} )$, since
$$
\left(\dst\frac{\partial}{\partial x^0} +
\dst\frac{\la + \la^{-1}}{2} \dst\frac{\partial}{\partial x^1}
- i \dst\frac{\la - \la^{-1}}{2} \dst\frac{\partial}{\partial x^2} \right)
\left( x^0 -  \dst\frac{\la + \bar{\la}}{1+\la \bar{\la}} x^1 +  i
\dst\frac{\la - \bar{\la}}{1+\la \bar{\la}}x^2 \right) = 0
$$

\subsection{Abelian Connection on 3-dimensional manifold.}
In the Abelian case we are given   a  1-form $\Omega=A_m(x)dx^m$, with
$\{A_m(x)\}$ taking values in  numbers and  fast decreasing as $x\rightarrow
\infty$,  defined up to an exact form.

\noindent {\bf Proposition 1.1}\newline
\begin{em}
Let us parametrize  light rays as follows:
$$
x=\al  t + \be,
$$
where $\al=(\al^0,\al^1,\al^2), \ \al_m \al^m =0$, $\al^0>0$ \quad
and $\al,\be \in {\bf R}^3$.
Let
$$
\varphi (\al, \be) =\dst\int_{x=\al t + \be}\Omega\equiv
\dst\int_{-\infty}^{+\infty} \al^i A_i (\al t + \be) dt.
$$
be the integral of 1-form over oriented light ray. Such integral gives us a
transformation from  1-forms $\Omega=A_m(x)dx^m$ to functions  on the space of
light rays $\varphi (\al, \be)$ :
$$
\varphi (\la \al, \be)=sgn\la \ \varphi (\al, \be)
$$
$$
\varphi (\al, \be + \la \al)=\varphi (\al, \be)
$$
\vskip 30pt

If we know $\varphi (\al, \be)$, we can reconstruct the Fourier
transform of the
1-form $\Omega$ outside the light cone, up to an exact form.
We will give two ways to do this:
\begin{enumerate}
\item
For pairs $\al, \eta$ such that
\bge
\bga
{(\al^1)}^2 + {(\al^2)}^2-{(\al^0)}^2 =0
\\[14pt]
\eta_0 \al^0 + \eta_1 \al^1 + \eta_2 \al^2 =0.
\ena
\ene
define
$$
\Phi(\al, \eta) = \dst\int_{{\bf R^3}/ \{ \al\}} \varphi (\al, \be) \
e^{i \be^m \eta_m} d_{\al} \be
$$
where the integral is over the factor space of ${\bf R^3}$
over the 1-dimensional subspace spanned by $\al$, and
$ d_{\al} \be =\vert \al^0 d\be^1\wedge d\be^2 +
 \al^1 d\be^2\wedge d\be^0+\al^2 d\be^0
\wedge d\be^1\vert$  is a nonoriented volume element on this space.

Using homogenious properties, we can choose
$$
\al^0=1
$$
For fixed $\eta$, (2) viewed as an equation for $\al$
has two real-valued solutions
$\al^{({\bf 1})}, \ \al^{({\bf 2})} $, provided
${\eta_1}^2 + {\eta_2}^2 -{\eta_0}^2 > 0$,
$$
\bga
\al^{({\bf 1})} = \left( 1 \ , \dst\frac{- \eta_0 \eta_1 - \eta_2
\sqrt{{\eta_1}^2 + {\eta_2}^2 - {\eta_0}^2} } {{\eta_1}^2 + {\eta_2}^2} , \
\dst\frac{ - \eta_0 \eta_2 + \eta_1 \sqrt{{\eta_1}^2 + {\eta_2}^2 - {\eta_0}^2}
} {{\eta_1}^2 + {\eta_2}^2}\right)
\\[20pt]
\al^{({\bf 2})} = \left( 1 \ , \dst\frac{ - \eta_0 \eta_1 + \eta_2
\sqrt{{\eta_1}^2 + {\eta_2}^2 - {\eta_0}^2} } {{\eta_1}^2 + {\eta_2}^2} , \
\dst\frac{ - \eta_0 \eta_2 - \eta_1 \sqrt{{\eta_1}^2 + {\eta_2}^2 - {\eta_0}^2}
} {{\eta_1}^2 + {\eta_2}^2}\right) ,
\ena
$$
and no real-valued solutions for ${\eta_1}^2 + {\eta_2}^2 -{\eta_0}^2 <
0.$\newline
For  $\eta$ such that ${\eta_1}^2 + {\eta_2}^2 - {\eta_0}^2>0$, we can
reconstruct
the 1-form up  to the exact form:
\bge
\bga
\eta_0 \hat{A}_1(\eta) - \eta_1\hat{A}_0(\eta)=
 \rho \left( \al^{({\bf 2}),2}\  \Phi(\al^{({\bf 1})}, \eta) - \al^{({\bf 1}),
2}\  \Phi(\al^{({\bf 2})}, \eta)\right)
\\[14pt]
\eta_1 \hat{A}_2(\eta) - \eta_2\hat{A}_1(\eta)=
 \rho \left( \al^{({\bf 2}),0}\  \Phi(\al^{({\bf 1})}, \eta) - \al^{({\bf 1}),
0}\  \Phi(\al^{({\bf 2})}, \eta)\right)
\\[14pt]
\eta_2 \hat{A}_0(\eta) - \eta_0\hat{A}_2(\eta)=
 \rho \left( \al^{({\bf 2}),1}\  \Phi(\al^{({\bf 1})}, \eta) - \al^{({\bf 1}),
1}\  \Phi(\al^{({\bf 2})}, \eta)\right)
\ena
\ene
where
$$
\rho =
\dst\frac{\eta_0}{\al^{({\bf 1}), 1} \al^{({\bf 2}), 2} - \al^{({\bf 1)}, 2}
\al^{({\bf 2}), 1}} =
\dst\frac{\eta_1}{\al^{({\bf 1}), 2} \al^{({\bf 2}), 0} - \al^{({\bf 1}), 0}
\al^{({\bf 2}), 2}} =
\dst\frac{\eta_2}{\al^{({\bf 1}), 0} \al^{({\bf 2}), 1} - \al^{({\bf 1}), 1}
\al^{({\bf 2}), 0}}
$$
\vskip 30pt
\item
There is another way to write the inversion formulas. Define
$$
\hat{\varphi} (\al, \eta)=\dst\int \varphi (\al, \be) \ e^{i \be^m \eta_m} d^3
\be
$$
to be the Fourier transform over all $\be$. Here $\al_m \al^m = 0$.\newline
We can reconstruct the connection up to the exact form as follows:\newline
Let us parametrise $\al$ by
$$
\al = (1, \cos\ga, \sin\ga).
$$
Then
$$
\bga
\dst\frac{{\eta_1}^2 + {\eta_2}^2}{2 {({\eta_1}^2 + {\eta_2}^2 -
{\eta_0}^2)}^{1/2}}\dst\int_{0}^{2\pi}(\eta_2 \cos\ga - \eta_1 \sin\ga) \
\hat{\varphi}(\al(\ga) , \eta) \dst\frac{d\ga}{2\pi}=
\\[10pt]
(\eta_2 \hat{A}_1(\eta) - \eta_1\hat{A}_2(\eta) ) \ \Theta({\eta_1}^2 +
{\eta_2}^2 - {\eta_0}^2)
\\[20pt]
\dst\frac{1}{2} {({\eta_1}^2 + {\eta_2}^2 - {\eta_0}^2)}^{1/2}
\dst\int_{0}^{2\pi}
 \hat{\varphi}(\al(\ga) ,\eta) \dst\frac{d\ga}{2\pi}=
\\[10pt]
(\hat{A}_0(\eta)- \dst\frac{\eta_0}{{\eta_1}^2 +
{\eta_2}^2}(\hat{A}_1(\eta)\eta_1+\hat{A}_2(\eta)\eta_2)\ \Theta({\eta_1}^2 +
{\eta_2}^2 - {\eta_0}^2)
\ena
$$
where  $$\Theta(x)= \cases{\ \ 1,& if $x>0$; \cr \ \ 0,&if $x<0$ \cr}$$.

\end{enumerate}
\end{em}
\vskip 20pt
We see that we can reconstruct the Fourier transform of the connection outside
the light cone, that is , for ${\eta_1}^2 + {\eta_2}^2 -{\eta_0}^2 > 0$, only.
To get the connection inside the cone, we need additional data, namely, we have
to use the complex-valued solutions of (2) for $\al$.

$A_m (x) dx^m$ is a 1-form, that is a linear function on vector fields $r_n
\dst\frac{\partial}{\partial x^n}$, with real $r_n$. Using linearity, we can
extend it to the linear function on vector fields $c_n
\dst\frac{\partial}{\partial x^n}$ with complex $c_n$. Using the 1-form
$A_m (x) dx^m$ as the connection,consider the paralel transport equation
\bge
\bga
\left(\dst\frac{\partial}{\partial x^0} +
\dst\frac{\la + \la^{-1}}{2} \dst\frac{\partial}{\partial x^1}
- i \dst\frac{\la - \la^{-1}}{2} \dst\frac{\partial}{\partial x^2} \right)
\mu (x,\la , \bar{\la}) =
\\[12pt]
\left(A_0  (x)+
 \dst\frac{\la + \la^{-1}}{2} A_1  (x)
- i \dst\frac{\la - \la^{-1}}{2} A_2  (x) \right)
\mu (x,\la ,\bar{\la})
\ena
\ene
Here $\la$ is a complex number,   $x \in {\bf R^3}$, $A_m (x)$ takes values
in numbers.  We have  parametrised the direction of the (complex) light ray by
$\al=(1, \dst\frac{\la + \la^{-1}}{2}, -i \dst\frac{\la - \la^{-1}}{2} )$.
\newline

Let us take the following solution of (4):
\bge
\mu (x,\la ,\bar{\la})=exp\left( \dst\int G(x-y,\la ,\bar{\la}) \left( 2 \la
A_0  (y)+
( {\la}^2 + 1) A_1  (y)
- i ({\la}^2 -1) A_2  (y) \right) d^3y \right),
\ene
where
\bge
\bga
G(x,\la,\bar{\la}) := \dst\int \dst\frac{e^{i\eta x}}{2\la \eta_0 + {\la}^2
(\eta_1 - i \eta_2) + (\eta_1 + i\eta_2)} \ \ \dst\frac{d^3 \eta}{i {(2\pi)}^3}
=
\\[20pt]
\dst\frac{\bar{\la} sgn(\la \bar{\la}- 1)}{\pi}\ \
\dst\frac{\delta (-x^0 (\la \bar{\la}+1) + x^1 (\la + \bar{\la}) -i x^2
(\la-\bar{\la}))}
{x^0 (\la \bar{\la}-1) + x^1 (\la -\bar{\la}) -i x^2 (\la+\bar{\la})}.
\ena
\ene

Here $$sgn(x)= \cases{\ \ 1,& if $x>0$; \cr -1,&if $x<0$ \cr}$$.

\noindent {\bf Lemma 1.1}\newline
\noindent {\em Define the functions of the light ray }  $S(x,\la,\bar{\la})$,
$T(x ,\ga)$ {\em as follows:}
$$
\bga
S (x,\la ,\bar{\la}) := \dst\frac{1}{\mu (x,\la, \bar{\la})}
\dst\frac{\partial}{\partial \bar{\la}} \mu (x,\la, \bar{\la})
\\[20pt]
T(x,\ga):=\dst\frac{\mu\left(x,\la,\bar{\la}\right)
\vert_{\la=(1-\varepsilon)e^{i\ga}}- \mu\left(x,\la,\bar{\la}\right)
\vert_{\la=(1+\varepsilon)e^{i\ga}} }{\mu\left(x,\la,\bar{\la}\right)
\vert_{\la=(1-\varepsilon)e^{i\ga}}}
\ena
$$
{\em where } $\mu (x,\la, \bar{\la})$ {\em is given by} (5), (6).
\vskip 20pt

\noindent {\em 1). If we are  given a connection, we can find}
$S(x,\la,\bar{\la})$, $T(x ,\ga),$ {\em which depend from 3 real parameters,
and given by }
$$
\bga
S (x,\la ,\bar{\la}) =  \dst\frac{sgn(\la \bar{\la}- 1)}{2\pi}\dst\int
{\delta}^{\ {\bf \prime}}\left( (\la \bar{\la}+1)(x^0 - y^0) - (\la +
\bar{\la})(x^1 - y^1) + i  (\la-\bar{\la}) (x^2 - y^2)\right)
\\[16pt]
 \left( 2 \la A_0  (y)+
( {\la}^2 + 1) A_1  (y)
- i ({\la}^2 -1) A_2  (y) \right)
\ena
$$
$$
\bga
T (x,\ga)= I - exp\left(\dst\frac{1}{2i\pi}\dst\int \dst\frac{\delta \left(-(x
^0 - y^0)+
(x^1 - y^1)\cos (\ga) + (x^2 - y^2)\sin (\ga)\right)}{(x^1-y^1)\sin
(\ga)-(x^2-y^2)\cos (\ga) }\right.
\\[14pt]
\left.  ( A_0 (y) + \cos(\ga ) A_1 (y) + \sin(\ga ) A_2 (y) ) \right)
\ena
$$
{The quantities} $S(x,\la,\bar{\la})$, $T(x ,\ga)$ {\em do not change in
the  light-like direction, }
$$
\bga
\left( \dst\frac{\partial}{\partial x^0} +
\dst\frac{\la + {\la}^{-1}}{2} \dst\frac{\partial}{\partial x^1}
- i \dst\frac{\la - {\la}^{-1}}{2} \dst\frac{\partial}{\partial x^2} \right)
S (x,\la , \bar{\la}) = 0,
\\[16pt]
\left( \dst\frac{\partial}{\partial x^0} +
\cos(\ga ) \dst\frac{\partial}{\partial x^1} +
\sin (\ga) \dst\frac{\partial}{\partial x^2} \right)
T (x,\ga ) = 0.
\ena
$$
{\em 2). If we are given }  $S(x,\la,\bar{\la})$ and  $T(x ,\ga)$, {\em we can
reconstruct from them the connection , up to a gauge transformation, using the
relations}
\bge
\bga
({\eta_1}^2 + {\eta_2}^2) \left(2\la\eta_0 + {\la}^2(\eta_1 - i\eta_2) + \eta_1
+ i\eta_2\right)\dst\int_{0}^{2\pi} t(\eta,\ga) \dst\frac{e^{i\ga}}{\la -
e^{i\ga}} \dst\frac{i d\ga}{2\pi}=
\\[16pt]
\left(2\la \left(A_0 ({\eta_1}^2 + {\eta_2}^2)- \eta_0(\eta_1 A_1 + \eta_2 A_2)
- i \eta_0(\eta_2 A_1 - \eta_1 A_2) \right) - 2i (\eta_2 A_1 - \eta_1
A_2)(\eta_1 + i \eta_2) \right)
\\[12pt]
\hskip 300pt \Theta({\vec{\eta}}^2 - {\eta_0}^2 ) ,
\\[32pt]
({\eta_1}^2 + {\eta_2}^2)\left(2\la\eta_0 + {\la}^2(\eta_1 - i\eta_2) + \eta_1
+ i\eta_2\right)\dst\int \hat{S}(\eta,\la_1,\bar{{\la}_1}) \dst\frac{1}{\la -
{\la}_1}\dst\frac{d{\la}_1 d{\bar{\la}}_1}{2\pi}=
\\[16pt]
\left(2\la \left(A_0 ({\eta_1}^2 + {\eta_2}^2)- \eta_0(\eta_1 A_1 + \eta_2 A_2)
- i \eta_0(\eta_2 A_1 - \eta_1 A_2)\right) - 2i (\eta_2 A_1 - \eta_1
A_2)(\eta_1 + i \eta_2) \right)
\\[12pt]
\hskip 300pt \Theta({\eta_0}^2 - {\vec{\eta}}^2).
\ena
\ene
Here
$$
t(\eta,\ga)= \dst\int ln\left(T(x,\ga)-1\right) e^{i\eta x}dx, \quad
\hat{S}(\eta,\la,\bar{\la}) = \dst\int S(x,\la,\bar{\la}) e^{i\eta\ga} dx.
$$

\vskip 20pt
{\it Proof.}\newline
\noindent Part 1)  can be checked directly, using the definitions. To prove 2),
notice that,
$$
\bga
\hat{S}(\eta,\la,\bar{\la}) = (2\pi) sgn(\la \bar{\la}- 1)\dst\frac{i\eta
_0}{{(1+\la \bar{\la})}^2} \delta\left(\eta_1 +\dst\frac{\la + \bar{\la}}{1+\la
\bar{\la}}\eta_0\right) \delta\left(\eta_2 - i \dst\frac{\la - \bar{\la}}{1+\la
\bar{\la}}\eta_0\right)
\\[16pt]
 \left( 2 \la {\af}_0  (\eta)+
( {\la}^2 + 1) {\af}_1  (\eta)
- i ({\la}^2 -1) {\af}_2  (\eta) \right)
\\[32pt]
t(\eta,\ga) := \hat{\left(ln\left( T-1\right)\right)}\ (\eta,\ga)= -4\pi
\delta\left(\eta_0 + \eta_1 \cos (\ga) + \eta_2 \sin (\ga)\right)
\\[16pt]
sgn\left(-\sin (\ga)\eta_1 + \cos (\ga)\eta_2\right)\left(\af_0 (\eta)+ \af_1
(\eta) \cos (\ga) + \af_2 (\eta)\sin (\ga)\right)
\ena
$$
After some computation, we  get (7).

Notice that the formulas (7) do not give the
connection at ${\eta_1}^2 + {\eta_2}^2 = 0$. We need to take ${\eta_1}^2 +
{\eta_2}^2 = \epsilon$ and go to the limit $\epsilon \rightarrow 0$, using
the fact that $A(\eta)$ is continuous. Thus, there are certain conditions on
the
spectral data, which were obtained from the 1-form, particularly at
${\eta_1}^2 + {\eta_2}^2 = 0$. We will not discuss these subtleties here.
\vskip 32pt

{\bf Remark}
The jump $T(x,\ga)$ is related to the integral of the connection over the real
light ray in the following way:
$$
\bga
\hat{\left(ln\left( T-1\right)\right)}\ (\eta,\ga) \ sgn\left(-\sin (\ga)\eta_1
+ \cos (\ga)\eta_2\right)=
\\[14pt]
-4\pi\delta\left(\eta_0 + \eta_1 \cos (\ga) + \eta_2 \sin (\ga)\right)
\left(\af_0 (\eta)+ \af_1 (\eta) \cos (\ga) + \af_2 (\eta)\sin (\ga)\right)= -2
\ffi,
\ena
$$
where $\ffi$ is the Fourier transform of $\ffa$ over $\be$, with
$\al=(1,\cos\ga,\sin\ga)$; $\ffa$ was defined above as the integral of 1-form
over the light ray.
\vskip 32pt

\subsection{Nonabelian Connection.}
In the nonabelian case we have the connection $A_{m} (x)dx^m$, $m=0,1,2$, over
real
3 dimensional space with Minkowski metric. Here
$A_{m}(x)$ are $n\times n$ complex matrices (we consider the $gl(n,C)$
connection only). Two connections $A_m (x)$ and
${\tilde{A}}_{m_1}(x)$, are related by a gauge transformation, iff
$$
A_m (x)= g(x){\tilde{A}}_{m}(x)g^{-1}(x)-g(x)\dst\frac{\partial}{\partial x^m}
g^{-1}(x)
$$
We assume that the connection is fast decreasing as $x\rightarrow \infty$.

For $\la \in C$ consider the equation ,
\bge
\bga
\left(2\la \dst\frac{\partial}{\partial x^0} +
( {\la}^2 + 1) \dst\frac{\partial}{\partial x^1}
- i ({\la}^2 -1) \dst\frac{\partial}{\partial x^2} \right)
\mu (x,\la , \bar{\la}) =
\\[12pt]
\left(2\la A_0  (x)+
 ({\la}^2 + 1) A_1  (x)
- i ({\la}^2 - 1) A_2  (x) \right)
\mu (x,\la ,\bar{\la}).
\ena
\ene
Here $\mu (x,\la ,\bar{\la})$ has values in $n\times n$ complex matrices
\vskip 16pt

{\bf Lemma 1.2}
\begin{em}
Choose the solution of (8) , given by the solution of the Fredholm integral
equation, (we assume that  this integral equation has the unique solution,
which is true at least for the connection of small norm):
\bge
\bga
\mu(x,\la,\bar{\la}) = I +
\\[14pt]
\dst\int  G(x-y,\la ,\bar{\la}) \left( 2 \la A_0  (y)+
( {\la}^2 + 1) A_1  (y)
- i ({\la}^2 -1) A_2  (y) \right)
\mu (y,\la) d^3 y,
\ena
\ene
where $ G(x,\la ,\bar{\la})$ is given by (6), and
$I$ is the unit matrix.

\noindent $\mu(x,\la,\bar{\la})$ , considered
as a funcion of $\la$ has a jump on the unit circle $ |\la |=1$.

\noindent Define $S (x,\la ,\bar{\la}), T(x,\ga)$ as follows:
\bge
S (x,\la ,\bar{\la})= \mu^{-1} (x,\la, \bar{\la})\dst\frac{\partial}{\partial
\bar{\la}} \mu (x,\la, \bar{\la}) ,\quad |\la|\neq 1;
\ene
\bge
\mu_+(x,\ga) - \mu_-(x,\ga)= \mu_+ (x,\ga) \left(I-T(x,\ga) \right),
\ene
where
$$
\mu_{\pm}(x,\ga) = \mu(x,\la,\bar{\la})\vert_{\la=(1\pm \varepsilon)e^{i\ga}}.
$$
Define also
$$
t_{\pm}(x,\ga)=\lim_{t\to -\infty}{\mu_{\pm}(\al t +\be,\ga)},
$$
is the asymptotic of $\mu_{\pm}$ at $t=-\infty$ along the light ray $x=\al t +
\be$, with $\al=(1,\cos\ga,\sin\ga)$
\vskip 25pt
\begin{enumerate}
\item For a given connection, spectral data $S(x,\la,\bar{\la})$, $T(x ,\ga)$
are well-defined, depend from 3 real parameters, and given by
\bge
\bga
S (x,\la ,\bar{\la}) =  \dst\frac{sgn(\la \bar{\la}- 1)}{2\pi}\dst\int
{\delta}^{\ {\bf \prime}}\left( (\la \bar{\la}+1)(x^0 - y^0) - (\la +
\bar{\la})(x^1 - y^1) + i  (\la-\bar{\la}) (x^2 - y^2)\right)
\\[16pt]
 \left( 2 \la A_0  (y)+
( {\la}^2 + 1) A_1  (y)
- i ({\la}^2 -1) A_2  (y) \right)
\mu (y,\la , \bar{\la}) d^3 y.
\ena
\ene
$$
T(x,\ga)= {(t_-(x,\ga)}^{-1}t_+(x,\ga),
$$
where
\bge
\bga
t_{+} (x,\ga)= I +\dst\frac{1}{i\pi}\dst\int \dst\frac{\delta \left(-(x ^0 -
y^0)+
(x^1 - y^1)\cos (\ga) + (x^2 - y^2)\sin (\ga)\right)}{(x^1-y^1)\sin
(\ga)-(x^2-y^2)\cos (\ga) +i0}
\\[14pt]
\left( A_0 (y) + \cos(\ga )+ A_1 (y)
\sin(\ga ) A_2 (y) \right)
\mu_+ (y,\ga ) d^3 y
\\[28pt]
t_{-} (x,\ga)= I -\dst\frac{1}{i\pi}\dst\int \dst\frac{\delta \left(-(x ^0 -
y^0)+
(x^1 - y^1)\cos (\ga) + (x^2 - y^2)\sin (\ga)\right)}{(x^1-y^1)\sin
(\ga)-(x^2-y^2)\cos (\ga) -i0}
\\[14pt]
\left( A_0 (y) + cos(\ga ) A_1 (y)+
sin(\ga ) A_2 (y) \right)
\mu_{-} (y,\ga ) d^3 y.
\ena
\ene
\item Spectral data $S(x,\la,\bar{\la})$, $T(x ,\ga)$ do not change along the
light ray,namely:
$$
\bga
\left(2\la \dst\frac{\partial}{\partial x^0} +
( {\la}^2 + 1) \dst\frac{\partial}{\partial x^1}
- i ({\la}^2 -1) \dst\frac{\partial}{\partial x^2} \right)
S (x,\la , \bar{\la}) = 0,
\\[16pt]
\left( \dst\frac{\partial}{\partial x^0} +
\cos(\ga ) \dst\frac{\partial}{\partial x^1} +
\sin (\ga) \dst\frac{\partial}{\partial x^2} \right)
T (x,\ga ) = 0.
\ena
$$
\item The spectral data $S(x,\la,\bar{\la})$, $T(x ,\ga)$ are invariant under
the gauge transformation
$$
A_m(x)\rightarrow g^{-1}(x) A_m (x) g(x) -
g^{-1}(x)\dst\frac{\partial}{\partial x^m}g (x)
$$
\end{enumerate}
\end{em}

{\it Proof}.In the proof of 1), we use the uniqueness of solution of (8). The
rest can be checked by a straitforward computation.
\vskip 30pt

Define the asymptotic $\mu(x,\infty$
\bge
\mu(x,\infty):=\lim_{\la \rightarrow \infty} \mu(x,\la,\bar{\la}).
\ene
where $\mu(x,\la,\bar{\la})$ is the solution of the integral eqution (8).

$\mu(x,\infty)$ can be obtained as the solution of the Fredholm integral
equation
\bge
\mu(x,\infty) = I + \dst\frac{1}{2\pi}\dst\int
\dst\frac{1}{x^1-y^1-i(x^2-y^2)}\delta(x^0-y^0)\Biggl(A_1 (y) -i A_2 (y)\Biggr)
\mu(y,\infty)d^3y .
\ene

{\bf Remark}\newline

1) As $\la \rightarrow \infty$ the equation (8) becomes
$$
\left(\dst\frac{\partial}{\partial x^1} - i \dst\frac{\partial}{\partial
x^2}\right) \mu(x,\infty) = \left( A_1 (x) -i A_2 (x) \right) \mu (x,\infty),
$$
or
$$
A_1 (x) -i A_2 (x)= \left((\dst\frac{\partial}{\partial x^1} - i
\dst\frac{\partial}{\partial x^2}) \mu(x,\infty)\right) {(\mu(x,\infty))}^{-1}
$$
Thus, for the case of $gl(n,C)$  $\mu(x,\infty)$ is the gauge where $A_1 (x) -i
A_2 (x)=0$.

2) For other Lie algebras , say for $so(n)$, in general it is not possible to
choose the
gauge $A_1 (x) -i A_2 (x)=0$. Thus, we need to add the normalization
$\mu(x,\infty)$ to the list of spectral data. $\mu(x,\infty)$ and $g(x)
\mu(x,\infty)$, where g(x) belong to the appropriate group, give rise to the
gauge equivalent connections. There will be certain conditions on spectral data
for the Lie algebras other than $gl(n,C)$. We will not investigate it here.
\vskip 14pt

{\bf Proposition 1.2}
If $S(x,\la,\bar{\la})$, $T(x ,\ga)$, and $\mu(x,\infty)$ are given, we can
reconstruct $\mu (x,\la, \bar{\la})$ as the solution of the Fredholm equation
\bge
\bga
\mu (x,\la, \bar{\la}) = \mu(x,\infty) + \dst\frac{1}{2\pi i} \dst\int
\dst\frac{1}{{\la}_1 -\la}\ \  \mu (x,{\la}_1, {\bar{\la}}_1) \ S(x,{\la}_1,
{\bar{\la}}_1) \ d{\la}_1 d{\bar{\la}}_1 +
\\[16pt]
\dst\frac{1}{2\pi } \dst\oint_{0}^{2\pi} \  \dst\frac{1}{e^{i\ga} -\la} \ \ \mu
(x,{\la}_1, {\bar{\la}}_1)\vert_{\la_1=(1-\varepsilon)e^{i\ga}} \left(
I-T(x,\ga) \right) \ e^{i\ga} d\ga,
\ena
\ene
\vskip 14pt
{\it Proof}\newline
This follows from the Cauchy formula for the $\bar{\partial}$ problem
$$
\dst\frac{\partial}{\partial \bar{\la}}\mu (x,\la, \bar{\la})= \mu (x,\la,
\bar{\la}) \ s(x,\la,\bar{\la})
$$
with $\mu (x,\la, \bar{\la})$ having the given jump at the unit circle and the
given asymptotic at $\la=\infty$
\vskip 14pt

\noindent {\bf Lemma1.3 (Inverse Transform)}\newline
\begin{em}
Suppose that we are given the spectral data  $s(x,\la,\bar{\la})$, $t_{\pm}(x
,\ga)$. From these spectral data, we can reconstruct     $A_m (x)$
up to the gauge transformation:

Let $\psi (x,\la, \bar{\la})$ be the solution of the integral Fredholm equation

\bge
\bga
\psi (x,\la, \bar{\la}) = I + \dst\frac{1}{2\pi i} \dst\int
\dst\frac{1}{{\la}_1 -\la}\ \  \psi (x,{\la}_1, {\bar{\la}}_1) \ S(x,{\la}_1,
{\bar{\la}}_1) \ d{\la}_1 d{\bar{\la}}_1 +
\\[16pt]
\dst\frac{1}{2\pi } \dst\oint_{0}^{2\pi} \  \dst\frac{1}{e^{i\ga} -\la} \ \
\psi (x,{\la}_1, {\bar{\la}}_1)\vert_{\la_1=(1-\varepsilon)e^{i\ga}} \left(
I-T(x,\ga) \right) \ e^{i\ga} d\ga,
\ena
\ene
There exist some $g(x)\in GL(n,C)$ such that connection is given by
\bge
\bga
g^{-1}(x)\left(A_1 (x) + i A_2 (x)\right)g(x) -   g^{-1}(x)
(\dst\frac{\partial}{\partial x^1}+ i \dst\frac{\partial}{\partial x^2}) g(x)=
\\[16pt]
\left(\Biggl(\Bigl( 2\la \dst\frac{\partial}{\partial x^0} +
( {\la}^2 + 1) \dst\frac{\partial}{\partial x^1}
- i ({\la}^2 -1) \dst\frac{\partial}{\partial x^2}\Bigr)\
\psi (x,\la , \bar{\la})\Biggr){\Biggl(\psi (x,\la ,
\bar{\la})\Biggr)}^{-1}\right) {\Biggr\vert}_{\la=0}
\\[50pt]
g^{-1}(x)\left(A_0 (x)\right)g(x) -   g^{-1}(x) \dst\frac{\partial}{\partial
x^0} g(x) = \dst\frac{1}{2}
\\[16pt]
\left(\dst\frac{\partial}{\partial \la}
\Biggl(\Bigl( 2\la \dst\frac{\partial}{\partial x^0} +
( {\la}^2 + 1) \dst\frac{\partial}{\partial x^1}
- i ({\la}^2 -1) \dst\frac{\partial}{\partial x^2}\Bigr)\
\psi (x,\la , \bar{\la})\Biggr){\Biggl(\psi (x,\la ,
\bar{\la})\Biggr)}^{-1}\right) {\Biggr\vert}_{\la=0}
\\[50pt]
g^{-1}(x)\left(A_1 (x) -i A_2 (x)\right)g(x) -   g^{-1}(x)
(\dst\frac{\partial}{\partial x^1}- i \dst\frac{\partial}{\partial x^2}) g(x)=0
\ena
\ene
\end{em}
\vskip 14pt

{\it Proof}\newline
Comparing with Prop. 1.2, we see that for $g(x)=\mu(x,\infty)$ (18) follows
from (16), (17).
\vskip 32pt

\section{Spectral data in dimension 4.}

\subsection{Abelian connection.}
\noindent {\bf Proposition 2.1}\newline
\begin{em}
Consider the light ray
$$
x=\al  t + \be,
$$
where $\al=(\al^0,\al^1,\al^2,\al^3), \ \al_m \al^m =0$, $\al^0>0$ \quad and
$\al,\be \in {\bf R}^4$.
Let
$$
\varphi (\al, \be) =\dst\int_{x=\al t + \be}\Omega\equiv
\dst\int_{-\infty}^{+\infty} \al^i A_i (\al t + \be) dt.
$$
be the integral of 1-form over oriented light ray. It is a function  on the
space of light rays .
$$
\varphi (\la \al, \be)=sgn\la \varphi (\al, \be)
$$
$$
\varphi (\al, \be + \la \al)=\varphi (\al, \be)
$$
The space of all light rays has dimension 5, therefore, there should be some
differential equation for $\varphi (\al, \be)$. This equation can be written as
\bge
\bga
{\left(\varepsilon^{iklm} n_i \dst\frac{\partial}{\partial \be^k} \al_l
\dst\frac{\partial}{\partial \al^m} \right)}^3 \ffa =
\\[8pt]
\qquad \left(n^p n_p
\dst\frac{\partial}{\partial \be^r} \dst\frac{\partial}{\partial \be_r} -
n^p n^r
\dst\frac{\partial}{\partial \be^p} \dst\frac{\partial}{\partial \be^r}\right)
{\left(\varepsilon^{iklm} n_i \dst\frac{\partial}{\partial \be^k} \al_l
\dst\frac{\partial}{\partial \al^m} \right)} \ffa
\ena
\ene
where $n^i$ is an arbitrary vector in  ${\bf R}^4$ , and
$\varepsilon^{iklm}$ is totally antisymmetric tensor with $\varepsilon^{0123}
=1$
\end{em}
\vskip 20pt

{\it Proof}\newline
This can be checked by the straightforward computation.
\vskip 25pt

\noindent {\bf Proposition 2.2}\newline
\begin{em}

If we know $\varphi (\al, \be)$, we can reconstruct the Fourier transform of
the
1-form $\Omega$ outside the light cone, up to an exact form. We will give two
ways to do this:
\begin{enumerate}
\item
For pairs $\al, \eta$ such that
\bge
\bga
{(\al^1)}^2 + {(\al^2)}^2 + {(\al^3)}^2-{(\al^0)}^2 =0
\\[14pt]
\eta_0 \al^0 + \eta_1 \al^1 + \eta_2 \al^2 +\eta_3 \al^3=0,
\ena
\ene
define
$$
\Phi(\al, \eta) =  \dst\int_{{\bf R^4}/ \{ \al\}} \varphi (\al, \be) \ e^{i
\eta_m \be^m }  d_{\al}\be
$$
to be the Fourier transform of the function $\varphi (\al, \be)$ over the
factor space $\be \in {\bf R^4}/\{ \al\}$. Here $ d_{\al} \be =\vert \al^0
d\be^1\wedge d\be^2\wedge d\be^3 + \al^1 d\be^2\wedge d\be^3 \wedge
d\be^0+\al^2 d\be^3 \wedge d\be^0 \wedge d\be^1 + \al^3 d\be^0\wedge
d\be^1\wedge d\be^2 \vert$ is a nonoriented volume element on this space. From
homogenious property, we can choose
$$
\al^0=1
$$
For fixed $\eta$, (20) viewed as an equation for $\al$, with $\al^0=1$,  has a
1-parametric family of real-valued solutions ,
 provided
$$
{\eta_1}^2 + {\eta_2}^2 - {\eta_0}^2 > 0:
$$
and no real-valued solutions for
$$
{\eta_1}^2 + {\eta_2}^2 + {\eta_3}^2- {\eta_0}^2 < 0:
$$

For  $\eta$ such that ${\eta_1}^2 + {\eta_2}^2 + {\eta_3}^2- {\eta_0}^2>0$, we
can reconstruct
the connection up  to the exact form. Let
$$
\al^{\bf (i)}, i=1,2,3.
$$
be any three solutions of (20), linearly independent as vectors in ${\bf R^4}$.
( Such 3 solutions always exist for ${\eta_1}^2 + {\eta_2}^2 - {\eta_0}^2>0$).)
Notice that for pairs $(\al,\eta)$, solving (20),
\bge
\Phi(\al, \eta)= \al^0 A_0 (\eta)+\al^1 A_1 (\eta)+\al^2 A_2 (\eta)+ \al^3
A_3(\eta).
\ene
Thus, from the linear algebra it follows that
$$
\bga
\eta_i \hat{A}_j(\eta) - \eta_j\hat{A}_i(\eta)=
 \dst\frac{\rho}{2} \dst\sum_{{k\neq i,j}\atop{l\neq i,j}} sgn (p(i,j,k,l))\
\\[20pt]
\left(
\left|
\matrix{
\al^{({\bf 2}),k}& \al^{({\bf 2}),l}\cr
\al^{({\bf 3}),k}& \al^{({\bf 3}),l}\cr}\right|
\ \Phi(\al^{({\bf 1})}, \eta)+
\left|
\matrix{
\al^{({\bf 3}),k}& \al^{({\bf 3}),l}\cr
\al^{({\bf 1}),k}& \al^{({\bf 1}),l}\cr}\right|
\ \Phi(\al^{({\bf 2})}, \eta)+
\left|
\matrix{
\al^{({\bf 1}),k}& \al^{({\bf 1}),l}\cr
\al^{({\bf 2}),k}& \al^{({\bf 2}),l}\cr}\right|
\ \Phi(\al^{({\bf 3})}, \eta)
\right)
\ena
$$
where, for each $p(i,j,k,l)$ to be a permutation of numbers $(0,1,2,3)$
$$
\rho =sgn(p(i,j,k,l))
\dst\frac{\eta_i}{\left|
\matrix{
\al^{({\bf 1}),j}& \al^{({\bf 1}),k}& \al^{({\bf 1}),l}\cr
\al^{({\bf 2}),j}& \al^{({\bf 2}),k}& \al^{({\bf 2}),l}\cr
\al^{({\bf 3}),j}& \al^{({\bf 3}),k}& \al^{({\bf 3}),l}\cr}
\right|}
$$
\vskip 30pt
\item
There is another way to write the inversion formulas. Define
$$
\hat{\varphi} (\al, \eta)=\dst\int \varphi (\al, \be) \ e^{i \be^m \eta_m} d^4
\be
$$
to be the Fourier transform over all $\be$. Here $\al_m \al^m = 0$.\newline
We can reconstruct the connection up to the exact form as follows:
Let
$$
\al=(1,\vec{\al}) , \ \vert \vec{\al} \vert = 1
$$
\bge
\bga
\vert\vec{\eta}\vert \dst\int \ffn \dst\frac{d \omega_\al}{{(2\pi)}^2} =
\left(A_0 - \dst\frac{\eta_0 \left(\vec{A} \vec{\eta}
\right)}{{\vec{\eta}}^2}\right)\Theta({\vec{\eta}}^2 - {\eta_0}^2)
\\[20pt]
\dst\frac{2{\vert\vec{\eta}\vert}^3}{{\eta}^2} \dst\int \ffn
\dst\sum_{\nu=1}^{3}\left(\delta_{\mu\nu}-\dst\frac{\eta_\mu
\eta_\nu}{{\vec{\eta}}^2}\right){\al}_\nu \dst\frac{d \omega_\al}{{(2\pi)}^2} =
\dst\sum_{\nu=1}^{3}\left(\delta_{\mu\nu}-\dst\frac{\eta_\mu
\eta_\nu}{{\vec{\eta}}^2}\right)A_\nu \ \  \Theta({\vec{\eta}}^2 - {\eta_0}^2)
\ena
\ene
where $\int d \omega_\al$ is the integral over the unit sphere.
\end{enumerate}
\end{em}
\vskip 20pt
We see that the formulas above give  the Fourier transform of the connection
inside the light cone only. To reconstruct the Fourier transform of the
connection inside the light cone, we need additional data. For the abelian
connection, there is a nice way to give supplementary data, using the
quaternions. It is not clear how to use the quaternions for the nonabelian
problem, and we will use different methods there.

\subsection{Supplementary Data for the Abelian Connection using Quaternions.}
Consider  purely imaginary quaternions
$$
q=q_1 i + q_2 j + q_3 k,
$$
where $q_1, q_2, q_3$ are real  numbers, and $i^2=j^2=k^2=-1$, $i j= -j i = k$
and so on. Define $\vert q \vert := {q_1}^2 +  {q_2}^2 +  {q_3}^2 $
For $q$ such that $\vert q \vert+\dst\frac{1}{\vert q \vert}< 6$ define the
'quaternionic light ray' $\al_{(q)}$:
\bge
\al_{(q)} =\dst\frac{1}{2}\Bigl( {(6 + q^2 +q^{-2})}^{\frac{1}{2}}, -i q+
q^{-1} i,   -j q+ q^{-1} j,  -k q+ q^{-1} k \Bigr)
\ene
For $q$ with $\vert q \vert=1$, \quad $\al_{(q)}$ is just the usual real  light
ray. For  $q$ with $\vert q \vert\neq 1$, $\al_{(q)}$ is a vector with values
in quaternions such that  ${\al_{(q) }}^m{\al_{(q)}}_m=0$.

{\bf Proposition}\newline
Suppose that in addition to $\Phi(\al, \eta)$, (21)  defined for pairs
$(\al,\eta)$ satisfying (20), with $\al$ real, we are given
\bge
\Phi(\al_{(q)}, \eta):= {\al_{(q)}}^m A_m (\eta),
\ene
which is defined for pairs $(\al_{(q) },\eta)$, with real $\eta$ and
quaternionic   $(\al_{(q) }$, (23) such that
\bge
{\al_{(q) }}^m \eta_m = 0
\ene
for  $\vert q \vert\neq 1$. Then we can reconstruct the Fourier transform of
the connection inside the light cone as well.
\vskip 16pt

\noindent Indeed, from (25) it follows that
$$
(q_1, q_2, q_3) = \rho (\eta_1,\eta_2,\eta_3)
$$
where $\rho$ should  be a real number, and
$$
\rho^2 \vec{\eta}^2=\dst\frac{3{\eta_0}^2-\vec{\eta}^2 \pm
\sqrt{8}|\eta_0|{({\eta_0}^2- \vec{\eta}^2)}^{\frac{1}{2}}}{{\eta_0}^2+
\vec{\eta}^2}
$$
For ${\eta_0}^2- \vec{\eta}^2>0$ such real $\rho$ always exist. Substituting in
$\Phi(\al_{(q)}, \eta)$ we have
$$
\bga
\Phi(\al_{(q)}, \eta)=\left(A_0 - \dst\frac{\eta_0 \left(\vec{A} \vec{\eta}
\right)}{{\vec{\eta}}^2}\right)
\dst\frac{{\vec{\eta}}^2 \rho}{\eta_0} (1-\dst\frac{1}{\rho^2 {\vec{\eta}}^2})
+
\\[14pt]
\rho  (1+ \dst\frac{1}{\rho^2 {\vec{\eta}}^2}) \Biggl((A_1\eta_2-A_1\eta_1) k +
(A_2\eta_3-A_3\eta_2) i + (A_3\eta_1-A_1\eta_3) j \Biggr)
\ena
$$
Thus, we can reconstruct $A(\eta)$ for ${\eta_0}^2- \vec{\eta}^2>0$ up to
an exact form.

\subsection{Nonabelian Connection}.
For the spectral data we need to use complex light rays. In dimension 4 is
highly nontrivial, since the
dimension of the space of all real light rays is 5, (that is higher than the
dimension of space-time). Also, the direction of the light-ray is
parametrized by the unit sphere  $S^2$, and there is no canonical way to
introduce the complex directions , thus we need to make choises. (It
would be nice to use quaternios, but in the nonlinear case it is not
yet clear how to write the inversion formula with quaternions). Thus
we accept a different approach. The main idea is to reduce the
4-dimensional case to the 3-dimensional, considered before.The
direction of the light ray is parametrized by a point on a
sphere. We will choose the
spherical coordinates for the direction of real light rays, and for
each fixed azimuthal angle we can go to complex direction in the
same way as we have done in the 3-dimensional case . Thus, we have made
the following choice  to go to complex light-rays:
\bge
\bga
\left(2\la \dst\frac{\partial}{\partial x^0} + \sin\Theta\left(
( {\la}^2 + 1) \ \dst\frac{\partial}{\partial x^1}
- i ({\la}^2 -1) \dst\frac{\partial}{\partial x^2}\right) + 2\la \cos\Theta
\dst\frac{\partial}{\partial x^3}\right)
\mu (x,\la , \bar{\la},\Theta) =
\\[12pt]
\left(2\la A_0  (x)
+ \sin\Theta\left( ({\la}^2 + 1) A_1  (x)
- i ({\la}^2 - 1) A_2  (x)\right)  + 2\la \cos\Theta A_3  (x)\right)
\mu (x,\la ,\bar{\la},\Theta)
\ena
\ene
{\bf Lemma 2.1}
\begin{em}
Choose the solution of (26) , given by the solution of the Fredholm integral
equation, (we assume that  this integral equation has the unique solution,
which is true at least for the connection of small norm):
\bge
\bga
\mu(x,\la,\bar{\la},\Theta) = I +
\dst\int  G(x-y,\la ,\bar{\la},\Theta)
\\[14pt]
\hspace{30pt}\left( 2 \la A_0  (y)+ \sin\Theta\Bigl(
( {\la}^2 + 1) A_1  (y)
- i ({\la}^2 -1) A_2  (y) \Bigr)  + 2 \la\cos\Theta A_3  (x)\right)
\mu (y,\la) d^3 y,
\ena
\ene
where $ G(x,\la ,\bar{\la},\Theta)$ is given by
\bge
\bga
G(x,\la,\bar{\la},\Theta) :=\dst\frac{\bar{\la} sgn(\la \bar{\la}- 1)}{\pi}\ \
\dst\frac{\delta \left(- (\la \bar{\la}+1) x^+ \sin\Theta  +  x^1 (\la +
\bar{\la}) -i
x^2 (\la-\bar{\la}) \right)\ \ \delta(x^-)}
{ \left((\la \bar{\la}-1) x^+  \sin\Theta  +  x^1 (\la -\bar{\la}) -i x^2
(\la+\bar{\la}) \right)}.
\ena
\ene
where $x^+ = \dst\frac{x^0 +\cos\Theta \ x^3}{1+ {\cos\Theta}^2}, x^- =
 -\cos\Theta \ x^0 + x^3$.

\noindent $\mu(x,\la,\bar{\la},\Theta)$ , considered
as a funcion of $\la$ has a jump on the unit circle $ |\la |=1$.

\noindent Define $S (x,\la ,\bar{\la},\Theta), T(x,\ga,\Theta)$ as follows:
\bge
S (x,\la ,\bar{\la},\Theta)= \mu^{-1} (x,\la, \bar{\la},\Theta)
\dst\frac{\partial}{\partial \bar{\la}} \mu (x,\la, \bar{\la},\Theta) ,\quad
|\la|\neq 1;
\ene
\bge
\mu_+(x,\ga,\Theta) - \mu_-(x,\ga,\Theta)= \mu_+ (x,\ga,\Theta)
\left(I-T(x,\ga,\Theta) \right),
\ene
where
$$
\mu_{\pm}(x,\ga,\Theta) = \mu(x,\la,\bar{\la},\Theta)
\vert_{\la=(1\pm \varepsilon)e^{i\ga}}.
$$
\vskip 20pt

1) The scattering data $S(x,\la ,\bar{\la},\Theta)$,
$T (x,\ga,\Theta)$  are given by
\bge
\bga
S (x,\la ,\bar{\la},\Theta) =
\dst\frac{sgn(\la \bar{\la}- 1)}{2\pi}
\\[14pt]
\dst\int {\delta}^{\ {\bf \prime}}\left( (\la \bar{\la}+1)\sin\Theta (x^+ -
y^+) -
(\la + \bar{\la})(x^1 - y^1) + i  (\la-\bar{\la}) (x^2 - y^2)\right) \
\delta(x^- - y^-) \\[14pt]
 \left( 2 \la A_0  (y)+
\sin\Theta\left( ( {\la}^2 + 1) A_1  (y)
- i ({\la}^2 -1) A_2  (y)\right)+ \cos\Theta A_3  (y)\right)
\mu (y,\la , \bar{\la},\Theta) d^4 y,
\\[20pt]
T(x,\ga,\Theta)= {\bigl(t_{-}(x,\ga,\Theta)\bigr)}^{-1} t_{+}(x,\ga,\Theta),
\ena
\ene
where
\bge
\bga
t_{+} (x,\ga,\Theta)= I +\dst\frac{1}{i\pi}\dst\int \dst\frac{\delta
\left(-\sin\Theta (x ^+ - y^+)+
 (x^1 - y^1)\cos (\ga) + (x^2 - y^2)\sin (\ga) \right)\
\delta(x^- - y^-)}{ (x^1-y^1)\sin (\ga)-(x^2-y^2)\cos (\ga)  +i0}
\\[14pt]
\left( A_0 (y) + \sin\Theta\left(\cos(\ga ) A_1 (y)
\sin(\ga ) A_2 (y)\right)+ A_3(y)\cos\Theta \right)
\mu_+ (y,\ga,\Theta ) d^4 y
\\[28pt]
t_{-} (x,\ga,\Theta)= I -\dst\frac{1}{i\pi}\dst\int \dst\frac{\delta
\left(-\sin\Theta (x ^+ - y^+)+
 (x^1 - y^1)\cos (\ga) + (x^2 - y^2)\sin (\ga)\right) \
\delta(x^- - y^-)}{ (x^1-y^1)\sin (\ga)-(x^2-y^2)\cos (\ga) -i0}
\\[14pt]
\left( A_0 (y) + \sin\Theta\left(\cos(\ga ) A_1 (y)
\sin(\ga ) A_2 (y)\right)+ A_3(y)\cos\Theta \right)
\mu_+ (y,\ga,\Theta ) d^4 y,
\ena
\ene

Scattering data   $S(x,\la ,\bar{\la},\Theta)$,
$T (x,\ga,\Theta)$  depend from 5 real parameters. They do not change along
the light-line:
$$
\bga
\left(2\la \dst\frac{\partial}{\partial x^0} +
\sin\Theta \left(( {\la}^2 + 1) \dst\frac{\partial}{\partial x^1}
- i ({\la}^2 -1) \dst\frac{\partial}{\partial x^2}\right) + 2\la \cos\Theta
\dst\frac{\partial}{\partial x^3}\right)
S (x,\la , \bar{\la},\Theta) = 0,
\\[14pt]
\left( \dst\frac{\partial}{\partial x^0} +
\sin\Theta \left(\cos(\ga ) \dst\frac{\partial}{\partial x^1} +
\sin (\ga) \dst\frac{\partial}{\partial x^2}\right) + \cos\Theta
\dst\frac{\partial}{\partial x^3} \right)
T(x,\ga,\Theta ) = 0.
\ena
$$
\vskip 20 pt

2)Scattering data  $S(x,\la ,\bar{\la},\Theta)$,
$T (x,\ga,\Theta)$ do not change under the gauge transformations
\end{em}

\vskip 22pt
The asymptotic as $\la\rightarrow\infty$ is given by
$$
\mu(x,\infty) = I + \dst\frac{1}{2\pi}\dst\int
\dst\frac{1}{x^1-y^1-i(x^2-y^2)}\delta(x^0-y^0)\delta(x^3-y^3)\left(A_1 (y) -i
A_2 (y\right) \mu(y,\infty)d^4y .
$$

{\bf Lemma}\newline
\begin{enumerate}
\item If we are given the scattering data  $S(x,\la ,\bar{\la},\Theta)$,
$T (x,\ga,\Theta)$, and we
know that the scattering data were obtained from some connection,
 we can reconstruct the connection $A(x)$ up to a
gauge  transformation, as follows:

Define $\psi (x,\la, \bar{\la},\Theta)$ to be the solution of the integral
Fredholm equation
\bge
\bga
\psi (x,\la, \bar{\la},\Theta) = I + \dst\frac{1}{2\pi i} \dst\int
\dst\frac{1}{{\la}_1 -\la}\ \  \psi (x,{\la}_1, {\bar{\la}}_1,\Theta) \
s(x,{\la}_1, {\bar{\la}}_1,\Theta) \ d{\la}_1 d{\bar{\la}}_1 +
\\[16pt]
\dst\frac{1}{2\pi } \dst\oint \  \dst\frac{1}{e^{i\ga} -\la} \ \ \psi
(x,{\la}_1, {\bar{\la}}_1,\Theta)\vert_{\la_1=(1-\varepsilon)e^{i\ga}} \left(
I-T(x,\ga,\Theta) \right) \ e^{i\ga} d\ga.
\ena
\ene
Define
\bge
\bga
W(x,\ga,\Theta):=\dst\frac{1}{2} e^{-i\ga}
\left(\left(\left(2\la \dst\frac{\partial}{\partial x^0} + \sin\Theta\left(
{\la}^2 ( \dst\frac{\partial}{\partial x^1}-i
\dst\frac{\partial}{\partial x^2}) + (\dst\frac{\partial}{\partial x^1}+i
\dst\frac{\partial}{\partial x^2})\right) +\right.\right.\right.
\\[14pt]
\hskip 110pt  \left.\left.\left.+2\la \cos\Theta \dst\frac{\partial}{\partial
x^3}\right)
\psi (x,\la , \bar{\la},\Theta) \right) \psi^{-1} (x,\la,
\bar{\la},\Theta)\right){\Biggr\vert}_{\la=(1-\varepsilon)e^{i\ga}}
\ena
\ene
Then there is some $g(x)\in GL(n,C)$ such that
\bge
\bga
g^{-1}(x)\Biggl( A_0  (x)
+ \sin\Theta\left( \cos\ga A_1  (x)+
\sin\ga A_2  (x)\right)  + \cos\Theta A_3  (x)\Biggr)
g(x) -
\\[14pt]
g^{-1}(x) \left(\dst\frac{\partial}{\partial x^0} + \sin\Theta\left(
\cos\ga \dst\frac{\partial}{\partial x^1} +\sin\ga \dst\frac{\partial}{\partial
x^2}\right) + \cos\Theta \dst\frac{\partial}{\partial x^3}\right)g(x)=
W(x,\ga,\Theta).
\ena
\ene
\item There
compatibility conditions  for the scattering data $s (x,\la
,\bar{\la},\Theta)$,
$t_{\pm} (x,\ga,\Theta)$ are the following :
$W(x,\ga,\Theta)$, which is a functional of the scattering data, defined by
(33),(34) should solve the equation
\bge
\bga
( u(\ga,\Theta) \dst\frac{\partial}{\partial \ga} + v(\ga,\Theta)
\dst\frac{\partial}{\partial \Theta}) \left(-\dst\frac{1}{\sin^2 \Theta}
\dst\frac{\partial^2}{\partial \ga^2} + \dst\frac{1}{\sin\Theta}(\sin\Theta
\dst\frac{\partial}{\partial\Theta})\right) W(x,\ga,\Theta)
=
\\[14pt]
2 ( u(\ga,\Theta) \dst\frac{\partial}{\partial \ga} + v(\ga,\Theta)
\dst\frac{\partial}{\partial \Theta})
 W(x,\ga,\Theta)
\ena
\ene
where  $u(\ga,\Theta) , \ v(\ga,\Theta) $ are arbitrary functions.
\end{enumerate}

\section{Discussion}
Consider the following pair of linear differential equations:
\bge
\bga\left\{
\bga
\left(2i\la (\dst\frac{\partial}{\partial  y_0}-A_0( y)) + \sin\Theta\left(
( {\la}^2 + 1) \ (\dst\frac{\partial}{\partial  y_1}-A_1( y))
- i ({\la}^2 -1) (\dst\frac{\partial}{\partial  y_2}-A_2( y))\right) +\right.
\\[10pt]
\left.+2\la \cos\Theta (\dst\frac{\partial}{\partial y_3}-A_3(y))\right)
\mu (x,\la , \bar{\la},\Theta) = 0
\\[30pt]
\left(
\Big( -({\la}^2 - 1) -\cos\Theta (\la^2+1)\Big)\ (\dst\frac{\partial}{\partial
y_1}-A_1(y)) + \Big(
 i ({\la}_2 -1)\cos\Theta +i (\la^2 + 1)\Big) (\dst\frac{\partial}{\partial
y_2}- \right.
\\[10pt]
\left. -A_2(y)) +2\la\sin\Theta (\dst\frac{\partial}{\partial
y_3}-A_3(y))\right)
\mu (y,\la , \bar{\la},\Theta) = 0
\ena\right.
\ena
\ene
The first equation in the pair coincides with the equation (5.9) at purely
imaginary time $x_0= - iy_0$

The compatibility conditions of the system (5.10) are the following:
\bge
F_{mn}(y) = \frac{1}{2} \varepsilon^{mnpq} F_{pq}(y),\\
\ene
where $F_{mn}(y)$ is the field strength,
\bge
F_{mn}(y)= -\dst\frac{\partial}{\partial y_m} A_n(y) +
\dst\frac{\partial}{\partial y_n} A_m(y) + [A_m(y), A_n(y)],
\ene
and $ \varepsilon^{mnpq}$ is the totally antisymmetric tensor with
$\varepsilon^{0123}=1$.
A connection $A(y)$ is  called a self-dual, if the field strength (5.12)
satisfies
equations (5.11). Thus, we have proved the following
\newtheorem{reym}{Remark}
\newtheorem{prym}{Proposition}
\begin{prym}
The compatibility condition of the system of equations (5.10) is that the
connection $A(y)$ is a self-dual connection.
\end{prym}
\vspace{15pt}
{\bf Remark 2. Light-Line Radon Transform and Crofton-Nambu Lagrangians.}
The Light-Line  Radon transform, as described in Chapter 4, gives a
transformation from a nonabelian connection on a 4 dimensional real Minkowski
space
to (matrix-valued) functions on the space of 2 dimensional planes in the 4
dimensional space.

Indeed, the functions on the space of such planes which do not contain a light
line  are present in our construction explicitly:
they correspond to the complex values of parameter, characterising the
direction of a light line, see (4.31). For planes which do contain a light
line, and therefore contain a family of parallel light lines, we can construct
a function of such planes
by integrating the function on the space of light lines over the family of
paralel light lines, contained in the plane.

\begin{prym}
Any function $\mu$ on the space of  two dimensional planes in the 4 dimensional
space
give rise to an action for a string theory  with the following
properties:\hfill\break
0)the action is reparametrization-invariant on a world sheet; its target is the
4 dimensional  Eucleadean space \hfill\break
1)for a world sheet which is topologically a plane, maps to 2 dimensional
planes in the 4 dimensional space are geodetic\hfill\break
2)if the function $\mu$ is positive, the "holomorphic" maps $X (\sigma - \tau)$
are geodetic, and, moreover, give a  surface with the minimal value of the
action.
\end{prym}
Let us parametrise a 2 dimensional plane in a 4 dimensional space $X \in {\Bbb
R^4}$ as follows:
\bge
\bga
\pmatrix{X_3\cr X_4\cr}= A \pmatrix{X_1\cr X_2\cr} +B,
\\[15pt]
A=\pmatrix{a_{31} & a_{32}\cr a_{41} & a_{42} \cr},\quad B=\pmatrix{b_3\cr
b_4\cr}
\ena
\ene
We will write a function on the space of 2 dimensonal planes in the 4
dimensional space as
$$
\mu(A,B)
$$
Then
 the string action for maps $X\equiv X(\sigma,\tau)$ can be written as
\bge
S = \dst\int \mu\left(A, \pmatrix{X_3\cr X_4\cr} - A \pmatrix{X_1\cr
X_2\cr}\right) \left|\det\pmatrix{
X_{1,\sigma} & X_{1,\tau} & 1 & 0\cr
X_{2,\sigma} & X_{2,\tau} & 0 & 1\cr
X_{3,\sigma} & X_{3,\tau} & a_{31} & a_{32}\cr
X_{4,\sigma} & X_{4,\tau} & a_{41}&  a_{42}\cr}\right| \ dA d\sigma d\tau
\ene
where $dA$ is an invariant measure on the Grassmanian.

A particular case of such action is the Nambu action, which is just the area of
the surface $X(\sigma,\tau)$

The Lagrangians as above were studied in \cite{croft}. In \cite{croft} such
Lagrangians were called Crofton Lagrangians, due to  a classical theorem of
Crofton, that
if we have a curve in two dimensions, and we count the number of intersection
points (without sign) of a curve with a line, and than average over all lines
with some measure, the result is proportional to the length of the curve.
Similar approach was used here.

Although it is not yet clear, how to use the Lagrangians above in the quantum
case,
the fact that Yang-Mills turns out to be related with a string theory through a
Radon transform is absolutely new and totally unexpected. It is a step towards
solving a problem to relate   Yang-Mills theory with
a string theory,  advocated by A.Polyakov.

{\bf Remark 3. Light-Line Radon transform, representation theory  of the group
$SL(2,R)$ in terms of integral geometry, and   $SL(2,R)$-type representation
theory without a group.}

Recall that the simpliest version of the Light-Line Radon transform, namely the
Light-Line Radon transfom of a function $f(x)$ on  a real Minkowski space of
dimension 3,  is given by the integral of the function $f(x)$ over all
light lines. We parametrise a light line as follows:
$$
\bga
x=\al t +\be, \\
\al=(1,\frac{\la+\la^{-1}}{2},\frac{\la-\la^{-1}}{2i}), \la=e^{i\ga}, 0\leq
\ga\leq 2\pi \\
\be=(0,\be_1,\be_2)
\ena
$$
Then the light line Radon transform $J$ of a function $f$,
$Jf\equiv\phi_f(\la,\be_1,\be_2)$ is given by
\bge
\bga
\phi_f(\la, \be_1,\be_2)=\dst\int_{-\infty}^{+\infty} f(t,
\frac{\la+\la^{-1}}{2} t +\be_1, \frac{\la-\la^{-1}}{2i} t + \be_2) dt\\
\ena
\ene

The formula (5.15) is similar to a formula in the representation theory of the
group $SL(2,R)$. Let me explain this point.
\vspace{15pt}

The group  $SL(2,R)$ is the group of real $2\times2$ matrices with the
determinant 1. The continuous series of unitary irreducible representations of
the group  $SL(2,R)$ is labelled by a multiplicative unitary character
$\pi (t)=|t|^{i\rho} {sgn}^\epsilon (t), \epsilon=0,1$. It can be realized on
the space of Schwarz class functions of one real variable $\varphi(x)$.
An operator $T_{\pi}(g)$ corresponding to the element $g \in G$, \quad
$g=\pmatrix{ g_{11} & g_{12} \cr
            g_{21} & g_{22} \cr}, \quad g_{11}g_{22}-g_{12}g_{21}=0 $ in
representation $\pi$ is given by:
$$
(T_\pi (g) \varphi)(x)=  \varphi\left(  \frac{g_{11} x+g_{21}}{g_{12} x+g_{22}
} \right)
\pi (g_{12} x+g_{22}) {|g_{12} x+g_{22} |}^{-1}.
$$
Thus,
$T_\pi (g)$ is an integral operator with kernel $K_\pi (g;\, x,y)$ given by a
generalized function
$$
K_\pi (g; x,y)=\delta \left( \frac{g_{11} x+g_{21}}{g_{12} x+g_{22} } - y
\right)
\pi (g_{12} x+g_{22}) {|g_{12} x+g_{22} |}^{-1}
$$
Let us now take a function on the group, $\Phi(g)$. To any such function
corresponds an operator with the kernel
$$
(K_\pi \Phi)(x,y) = \dst\int K_\pi (g;\, x,y) \Phi(g) dg,
$$
where $dg=\dst\frac{dg_{12}dg_{22}dg_{21}}{|g_{22}|}$ is an invariant measure
on the group.
Let us choose the coordinates on the group as follows:
$$
a_1=g_{12}, a_2=g_{22}, a_3=g_{11},
$$
and let us write the functuion on the group as $\Phi(a_1,a_2,a_3)$. Then it is
easy to check that
$$
(K_\pi \Phi)(x,y) = \dst\int \Phi(t, \la-xt, \la^{-1} + yt) \pi(\la) \la^{-1}
d\la dt
$$
If we know $(K_\pi \Phi)(x,y)$ for all unitary multiplicative characters on the
real line $\pi$, we know also
\bge
L(\la, x, y)= \dst\int_{-\infty}^{\infty} \Phi(t, \la +xt, \la^{-1} + yt)  dt
,\la\in {\Bbb R},
\ene
using the Mellin transform.

Compare this with the formula for the light-line Radon Transform.
$$
\bga
\phi_f(\la, \be_1,\be_2)=\dst\int_{-\infty}^{+\infty} f(t,
\frac{\la+\la^{-1}}{2} t +\be_1, \frac{\la-\la^{-1}}{2i} t + \be_2) dt,
\la=e^{i\ga}, 0\leq\ga\leq 2\pi
\ena
$$
Let us introduce the function $F$ as follows:
$$
f(x_1,x_2,x_3) ={x_1}^2
F\left(\frac{1}{x_1},\frac{x_2}{x_1},\frac{x_3}{x_1}\right).
$$
Then
\bge
\bga
\phi_f(\la, \be_1,\be_2)= -\dst\int_{-\infty}^{+\infty} F(t,
\frac{\la+\la^{-1}}{2} +\be_1t, \frac{\la-\la^{-1}}{2i}  + \be_2 t) dt,
\la=e^{i\ga}, 0\leq\ga\leq 2\pi
\ena
\ene

Thus, for the group $SL(2,R)$, we have integrals of a function of 3 real
variables over all lines
intersecting a hyperbola $a_1=0, a_2 a_3=1$, and for a light-line Radon
transform, we have integrals of a function of 3 real variables over all lines
intersecting a circle $x_1=0, {x_2}^2 + {x_3}^2 =1$

In representation theory of  the group $SL(2,R)$, there are Plancherel-type
formulas, which allow to reconstruct a function on the group from the kernels
of operators of its unitary irreducible representations. The representation
theory without
a group is the problem to reconstruct a function from the integrals of the
function over lines intersecting an arbitrary algebraic curve (this problem
makes sence in any dimension $n=2,3,4,\ldots .$ The idea to have a
representation theory without a group is due to I.Gelfand. For a large class of
curves, this
problem was solved in \cite{ggz1}. The light-line Radon Transform, described
above,  is, in
fact,  a nonlinear version of the problem of a representation theory without a
group.

\newpage
{\bf Acknowledgements}
The author is grateful to E.Witten for suggesting the
problem and discussions. This work is part of the
project with   I.Gelfand and M.Graev  in integral geometry and
its applications. Discussions with A. Fokas, A. Migdal, S. Shenker, A.
Zamolodchikov were
very valuable. Graduate Fellowship from the High Energy
Theory group at Rutgers Univ. is appreciated .

\end{document}